\documentclass[preprint,authoryear,12pt]{elsarticle}

\usepackage{a4wide}
\usepackage[english]{babel}
\usepackage{amssymb}
\usepackage{amsmath}
\newcommand{\dafg}{\textrm{d}}
\newcommand{\tol}{\texttt{tol}}
\newcommand{\Meen}{\mbox{\textsc{Method 1}}}
\newcommand{\Mtwee}{\mbox{\textsc{Method 2}}}
\newcommand{\Mdrie}{\mbox{\textsc{Method 3}}}
\newcommand{\Minit}{\mbox{\textsc{InitMan}}}
\newcommand{\MeenS}{\mbox{\textsc{Method 1}} }
\newcommand{\MtweeS}{\mbox{\textsc{Method 2}} }
\newcommand{\MdrieS}{\mbox{\textsc{Method 3}} }
\newcommand{\MinitS}{\mbox{\textsc{InitMan}} }

\usepackage{graphicx}

\journal{Computer Methods in Applied Mechanics and Engineering}

\begin{document}

\begin{frontmatter}



\title{An approach for both the computation of coarse-scale steady state solutions and initialization on
a slow manifold}


\author[label1,label4]{Christophe Vandekerckhove}
\author[label6,label2]{Benjamin Sonday}
\author[label3]{Alexei Makeev}
\author[label4]{Dirk Roose}
\author[label2,label5]{Ioannis G. Kevrekidis}

\fntext[label1]{Current address:  Markt 62, 9800 Deinze, Belgium}
\fntext[label6]{Corresponding author:  bsonday@math.princeton.edu}
\address[label4]{Department of Computer Science, K.U. Leuven, B-3001 Heverlee, Belgium}
\address[label2]{Program in Applied and Computational Mathematics, Princeton University, Princeton, NJ 08544}
\address[label3]{Moscow State University, Faculty of Computational Mathematics and Cybernetics (BMK), Moscow, 119899, Russia}
\address[label5]{Department of Chemical Engineering, Princeton University, Princeton, NJ 08544, USA}

\begin{abstract}
We present a simple technique for the computation of coarse-scale steady states of dynamical systems with
time scale separation in the form of a ``wrapper" around a fine-scale simulator.
We discuss how this approach alleviates certain problems encountered by comparable existing approaches, and illustrate its
use by computing coarse-scale steady states of a lattice Boltzmann fine scale code.
Interestingly, in the same context of multiple time scale problems, the approach can be slightly modified to
provide initial conditions {\it on the slow manifold} with prescribed coarse-scale observables.
The approach is based on appropriately designed short bursts of the fine-scale simulator whose results are used to track changes in the coarse variables of interest, a core component of the equation-free framework.

\end{abstract}

\begin{keyword}
slow manifold \sep coarse-graining \sep time-stepper \sep initialization

\end{keyword}

\end{frontmatter}


\section{Introduction} \label{intro}

For many problems in science and engineering, the best available model is given
at a  fine-scale level, while we would like to analyze its behavior at a much
coarser-scale ``system level".
To bridge the gap between the scale of
the available model and the scale of interest, one typically attempts to
derive a reduced model in terms of an appropriate set of variables (the
``coarse observables'').
In many cases, the derivation of such a reduced model hinges on the
existence of a low-dimensional, attracting, invariant slow manifold, which can
be parametrized in terms of these observables.
In fine-scale simulations, all
initial conditions are quickly attracted towards this slow manifold, on which
the reduced dynamics subsequently evolve.
In other words, the
remaining fine-scale model variables quickly become functionals of (become ``slaved
to'') the observables, and  the fine-scale model state can be accurately
described in terms of the observables only.

Although it should in principle always be possible to derive a reduced
dynamical model for a system possessing a slow manifold (see, e.g., \cite{qssa} and \cite{pea}),
one may fail to do
so, for instance because the closures required in its construction are not
accurate or even unavailable. 
For this class of problems, Kevrekidis and coworkers proposed the
equation-free framework (\cite{ecmce}),
which allows one to perform coarse-scale computations
based on appropriately initialized short fine-scale simulations.
(The term ``equation-free'' emphasizes that the derivation of reduced equations
 is circumvented.)
 In this paper, we propose and study two related equation-free algorithms that are designed to
(1) compute stable or unstable coarse-scale steady state solutions, or to (2)
systematically initialize a fine-scale model with prescribed values of the coarse observables (see, e.g., prior work in \cite{csp,ildm,aim,cmalc,ptasm,aenio,curry,lorenz} and some further mathematical analysis in \cite{kreiss,kaper}).
As we will see below, the gain in efficiency obtained with these algorithms
stems from the fact that the bulk of the
computations are only performed in the space of the observables, which is
typically low-dimensional compared to the full fine-scale variables space.

The outline of this paper is as follows:
In Section \ref{STST}, we address the computation of coarse-scale steady state solutions.
We succinctly illustrate certain problems encountered by existing, time-stepper
based algorithms, and propose a modification that alleviates these problems.
Section \ref{INITI} then discusses a variation of the latter approach that helps
initialize the fine-scale simulator {\it on the slow manifold}.
Section \ref{numerics} uses a lattice Boltzmann fine scale simulator to demonstrate
the application of both algorithms; the error characteristics of the algorithm
is presented in Section \ref{err_numerics} and we conclude with a brief discussion.

\section{Computing coarse-scale steady state solutions}
\label{STST}
In this section, we outline two existing methods to compute
coarse-scale steady state solutions, as well as the modification leading
to the  proposed, third method, with the help of a simple model problem.
Specifically, we consider the two-dimensional linear
 time integrator
\begin{equation}
\label{stepper} \left[ \begin{array}{c}
x  \\
y  \\
\end{array} \right]_{n+1}= \left[ \begin{array}{c c}
\cos(\alpha) & \cos(\beta)  \\
\sin(\alpha) & \sin(\beta)  \\
\end{array} \right]
 \left[ \begin{array}{c c}
0.999 & 0  \\
0 & 0.1  \\
\end{array} \right]
\left[ \begin{array}{c c}
\cos(\alpha) & \cos(\beta)  \\
\sin(\alpha) & \sin(\beta)  \\
\end{array} \right]^{-1}
\left[ \begin{array}{c}
x  \\
y  \\
\end{array} \right]_n
\end{equation}
as our fine-scale simulator with $x$ being the  ``coarse observable".
This arises, for instance, as the explicit Euler integrator for the system
\begin{equation}
\frac{d}{dt}\left[\begin{array}{c} x \\ y \end{array} \right] = M\,\left[\begin{array}{c} x \\ y \end{array} \right]
\end{equation}
with time step $\Delta\,t$, where
\begin{equation}
\Delta t\, M = \left[ \begin{array}{c c}
\cos(\alpha) & \cos(\beta)  \\
\sin(\alpha) & \sin(\beta)  \\
\end{array} \right]
 \left[ \begin{array}{c c}
-0.001 & 0  \\
0 & -0.9  \\
\end{array} \right]
\left[ \begin{array}{c c}
\cos(\alpha) & \cos(\beta)  \\
\sin(\alpha) & \sin(\beta)  \\
\end{array} \right]^{-1}.
\end{equation}
The slow manifold is then $y=\tan(\alpha) x$; note that this manifold can be parametrized by (is the graph of
a function over) our chosen observable as long as  $\alpha \not = \pm \pi/2$.
Any fine-scale
initial condition $(x,y)$ is quickly attracted to this slow manifold along a
trajectory that (at least far enough away from the slow manifold)  approximates a
line with slope $\tan(\beta)$.

\subsection{Description of the methods} \label{descr}
The obvious fine-scale steady state solution of \eqref{stepper} is $(x^*,y^*)=(0,0)$,
and this is a stable steady state.
In this section, we initially illustrate two existing methods whose purpose is to  approximate
the {\it coarse-scale} steady state solution $x^*=0$.
These two methods
are based directly on the concept of the {\it coarse time-stepper} (\cite{csaba2,ecmce}).
They approximate the value of $x^*$ by solving
\begin{equation}
\label{FPF12}
 \Phi(x,\tau)-x=0,
\end{equation}
where $\Phi(x,\tau)$ denotes a coarse time step over time $\tau$ with initial condition $x$.
Each such coarse time step (of time $\tau$) consists of the following three substeps:
\begin{enumerate}
 \item \emph{lifting}, in which an appropriate fine-scale state is constructed
according to the value of the observable $x$. \item \emph{simulation}, in which
the fine-scale state is evolved over time $\tau$, with $\tau$ large enough to allow
 $y$ to get slaved to  $x$, but small compared to the
 coarse time scales. 
 \item
\emph{restriction}, in which  the value of the observable is extracted from the
resulting fine-scale state. \end{enumerate}

If the value of $x$ \emph{does not change} substantially during the fast transients
towards the slow manifold, it may suffice to choose arbitrary (say, consistently the same)
values for the remaining fine-scale
variables (in this case, just $y$) in the lifting step.
This type of lifting will from now on
be called \emph{arbitrary lifting}, and the resulting method for approximating
the coarse-scale steady state solution will be called \Meen.

If the value of $x$ \emph{does change} substantially during the fast
transients towards the slow manifold, a more accurate initialization (i.e., an
initialization closer to the slow manifold) should be used.
In \cite{cmalc,ptasm}, it was
shown that an accurate initialization can be obtained with the so-called
\emph{constrained runs
lifting} scheme.
In its simplest form, this scheme determines, for a value $x^0$ of the observable,
the value of $y$
so that \mbox{$\textrm{d} y (x^0,y)/ \textrm{d}t  =0$}.
The intuitive reason why this condition (or, more generally, a condition demanding that
the $y$-time derivative to be bounded) yields a state $(x^0,y)$ close to
the slow manifold is that time differentiation amplifies fast components more
than slow components, so that, if the time derivatives are small, the
fast components in the remaining fine-scale variables are small.
In \cite{ptasm} it was rigorously shown that, under certain conditions,
the resulting state is indeed a good approximation to the desired point on the
slow manifold.
In practice, it is often convenient to  approximate the
derivative  $\textrm{d} y(x^0,y) / \textrm{d}t$  numerically, e.g. using forward  differences.
In
\cite{cmalc,ptasm} it was shown that a functional iteration can
then, in many cases, be used to find the zero of the resulting forward
difference condition.
If the step size of the
functional iteration and the forward difference formula are
both equal to the step size of the fine-scale simulator $\Delta t$,
the functional iteration takes the following form:
\begin{enumerate}
\item[0.] Initialize $y$ as well as possible. Then
start the iteration 1--3.
\item[1.] Evolve the fine-scale simulator over one time step of size
$\Delta t$, starting from $(x^0,y)$. 
\item[2.] Reset the value of $x$ to its original value $x^0$.
 \item[3.] If the difference between the current and the previous value of $y$
 is smaller than a certain tolerance \tol: end the iteration.
Else: go to 1. \end{enumerate}
The  iterative scheme above  will further be called
the \emph{constrained runs functional iteration} (abbreviated CRFI), and the method for
approximating the coarse-scale steady state solution that consists of solving
equation (\ref{FPF12}) with $y$ systematically initialized this way will be
called \Mtwee.

In some cases, \MeenS may produce very
inaccurate results when the value of $x$ changes substantially during
the fast transients toward the slow manifold.
\Mtwee, on the other hand, may find an accurate approximation to the exact
solution $x^*$ (more precisely, the error can be made arbitrarily small by decreasing the value of $\tol$).
%
This can easily be seen from the fact that
\begin{itemize}
 \item $\dafg x / \dafg t$ (or its finite difference approximation) is zero in
the coarse-scale steady state solution (this is just equation (\ref{FPF12})).
 \item $\dafg y / \dafg t$ (or its finite difference approximation) is zero at
the fixed point of the constrained runs functional iteration (because of the nature of this iteration).
 \end{itemize}
In essence, \MtweeS
computes the fine-scale steady state solution as a ``splitting scheme", by solving the
system  $\dafg y / \dafg t=0$ within each step of an outer solver for the system
$\dafg x / \dafg t=0$.
As a result, the computational complexity of \MtweeS
may be as large as that of directly solving the full fine-scale model (which is exactly
what we wanted to avoid).
%
%
Moreover, \MtweeS may fail when the constrained runs functional iteration {\it does not
converge} to the correct fine-scale state near the slow manifold (the iteration
may actually be unstable, or it may converge to a solution that does not
correspond to a state close to the slow manifold).
In some cases, these
convergence issues may be overcome by using a more advanced computational
approach such as the one presented in \cite{aenio}, yet then the issue of
overall computational complexity also remains.


To cope with the potential accuracy,  convergence or efficiency issues, we now propose a
third method, \Mdrie, to compute coarse-scale steady state solutions.
Instead of solving
\eqref{FPF12}, this method solves \begin{equation} \label{FPF3}
\Phi(x,\tau+\tau')-\Phi(x,\tau)=0, \end{equation} in which $\Phi(x,\tau)$, $\Phi(x,\tau+\tau')$
denote coarse time-steps over times $\tau$, $\tau+\tau'$ in which we use the arbitrary
lifting scheme.
As before, the value of $\tau$ should be large enough to allow
$y$ to get slaved to $x$, but small compared to the
coarse time scales.
The variable $\tau'$ represents the time interval
of an additional simulation on (or very close to) the slow manifold.
If $\overline{x}$ denotes
the solution of \eqref{FPF3}, we  expect
$\Phi(\overline{x},\tau)=\Phi(\overline{x},\tau+\tau')$ (and \emph{not} $\overline{x}$) to be a very good
approximation to the exact coarse-scale steady state $x^*$, as the finite
difference approximation of $\dafg x / \dafg t$ based on two points on the slow manifold is then
zero.
The obvious advantages compared to \MtweeS are that this method is
conceptually simpler, that it does not involve the (potentially unstable)
constrained runs functional iteration, and that we are no longer solving any
systems of the form \mbox{$\textrm{d} y (x^0,y)/ \textrm{d}t  =0$}
in the space of the remaining (fine scale) variables (which may be an advantage when
that space is large compared to the space of the observables $y$).

%

To summarize: we have outlined three different methods to find coarse-scale
steady state solutions.
In \Meen, we solve \eqref{FPF12}, in which
$\Phi(x,\tau)$ represents a coarse time step  based on arbitrary lifting.
In
\Mtwee, we solve \eqref{FPF12}, in which $\Phi(x,\tau)$ represents a coarse
time step based on the constrained runs functional iteration.
In \Mdrie, we
solve \eqref{FPF3}, in which $\Phi(x,\tau)$ represents a coarse time step
based on arbitrary lifting.

\subsection{Numerical illustration}
We now illustrate the performance of the three methods described above
 for the model problem \eqref{stepper} with $\alpha=\pi/6$ and
$\beta = -\pi/3$.
As the smallest eigenvalue of the Jacobian matrix of the
time integrator \eqref{stepper} is $\lambda_{2}=0.1$, it takes about 16 time steps
for an $\mathcal{O}(1)$ initial condition to reach the slow manifold
$y=x/\sqrt{3}$ up to machine precision.

\begin{figure}
\begin{center}
\includegraphics[width=150mm]{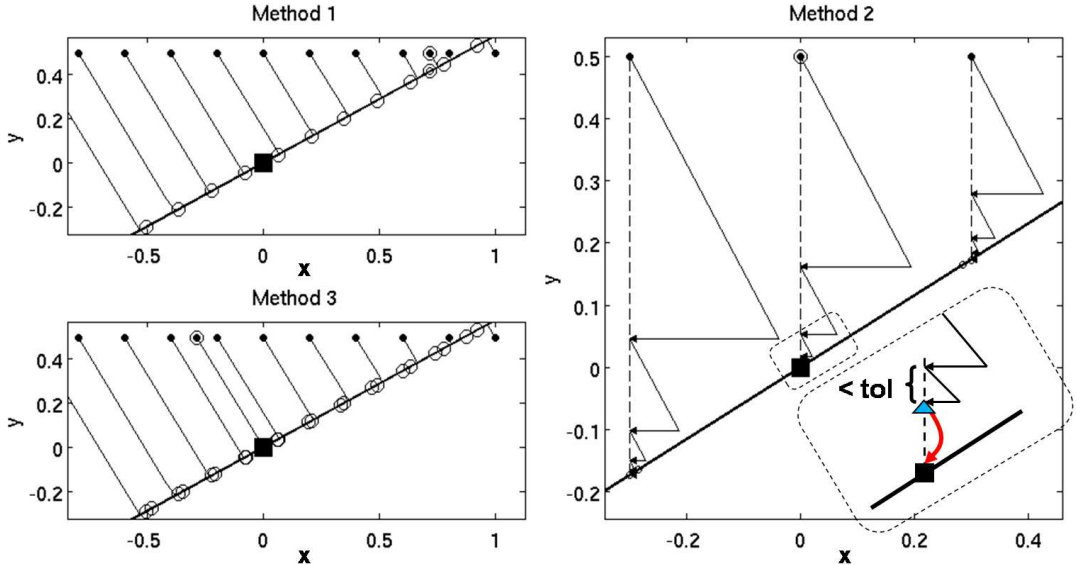}
\caption{An illustration of the three methods used to compute coarse-scale steady states.  Shown
are the slow manifold, fast transients, and computed coarse-scale steady states.  See
text for further explanation.\label{Fig1}}
\end{center}
\end{figure}

Some examples of a single iteration with \MeenS are given in Figure \ref{Fig1} (top-left).
The slow
manifold is represented by the thick line; the full model steady state solution
$(x^*,y^*)=(0,0)$ on the manifold is indicated by the filled square.
For various initial values of $x$, indicated by the filled circles, we perform a
 time step with a coarse time-stepper that is based on the arbitrary
lifting scheme $x \mapsto (x,1/2)$ and with $\tau=50 \Delta t$
(remember that $\Delta t$ is the time step of the fine scale time-stepper
\eqref{stepper}).
The fine-scale trajectories are represented by the fine
lines; note that the slope of these trajectories is approximately
$\tan(-\pi/3)=-\sqrt{3}$.
The end points of the trajectories are
indicated by the open circles; these points clearly lie on the slow manifold.
The solution of \Meen, the $x$-coordinate of the small open circle, is also shown; the initial condition of the
corresponding simulation trajectory is encircled as well.  For this trajectory, the
$x$-values of the initial and end points are equal, as demanded by \eqref{FPF12}.
In this case, however, the value of $x$ found is $0.719$, which is clearly
different from the exact value $x^*=0$.
As mentioned above, this is due to the fact that the value of $x$ changes
substantially during the fast transients towards the slow manifold ($\beta \not
\approx \pm \pi/2$); clearly, the solution found with \MeenS depends on the value of $\tau$.

An illustration of \MtweeS is given in Figure \ref{Fig1} (right).
Again, the
slow manifold is represented by the thick line and the full model steady
state solution $(x^*,y^*)=(0,0)$ is indicated by the filled square.
From
various initial values of $x$, indicated by the filled circles, we perform
 the constrained runs functional iteration with $\tol =10^{-15}$, after which we
perform an additional simulation over the time interval  $\tau=50 \Delta t$
(here, we only chose such a large value of $\tau$ to make Figure 1 more clear;
in practice there is little reason to use such a large value of $\tau$).
The
simulation trajectories and the resetting of the observable in the
constrained runs functional iteration are represented by the fine lines and arrows, respectively.
The end points of the constrained runs
functional iteration and the simulation trajectory afterwards are indicated by
the open circles.
We  observe that, in this example, the constrained runs
functional iteration always brings us very close to the slow manifold.
The final point reached by \MtweeS is also shown; the initial condition of the corresponding
simulation trajectory is encircled.
As in the case of \Meen, the
initial value of $x$ is  the same as the end value of $x$, as demanded by \eqref{FPF12}.
In this case, however, the value of $x$ found is $3.02
\cdot 10^{-15}$, which, as expected, approximates the exact value
$x^*=0$ up to the tolerance $\tol$.
In the lower right of this figure, one can see the details of \Mtwee:  first, the constrained runs
functional iteration is performed until the difference in successive $y$ values is less than $\tol$,
leaving us at the (blue) triangle; next, we perform a coarse time step of length $\tau = 50 \Delta t$
using the $y$-coordinates of this triangle for the \textit{lifting}, shown as the curvy (red) arrow; finally, because the $x$-coordinate
of the (blue) triangle coincides with the head of the curvy (red) arrow (meaning that $\Phi(x,\tau)-x=0$), we determine
that we are at a coarse-scale steady state.

Due to the fact that the constrained runs lifting brings us very close to the manifold,
we may even use a smaller value of $\tau$ than
in \Meen; even if $\tau=\Delta t$, we
obtain $x=1.05 \cdot 10^{-13}$.
%
It is also worth mentioning that if we had chosen $\beta=\pi/4$,
the constrained runs functional iteration would not have converged, as the
iteration is then unstable.
For our model problem, this can easily be rationalized
using geometric arguments (\cite{ch_thesis}).

An illustration of \MdrieS is given in Figure \ref{Fig1} (bottom-left).
Again, we used the same line styles and markers as before, but now
both the end points of the simulation trajectories over time
$\tau=50 \Delta t$ and $\tau+\tau'=100\Delta t$ are indicated by the open circles
(as before, we only chose such a large value of $\tau'$ to make Figure 1 more clear;
in practice there is little reason to use such a large value of $\tau'$).
The solution of \MdrieS is
also shown; the initial condition and the end point of the corresponding
simulation trajectory are encircled.
For this solution, the
initial value of $x$ {\it is not the same} as the end value of $x$; yet the
$x$-values of the end points of the simulation trajectories after  the time
interval $\tau$ and $\tau+\tau'$  coincide, as demanded for
in \eqref{FPF3}.
The value of $x$ found is now $1.89 \cdot 10^{-18}$, which
corresponds to the exact value $x^*=0$ up to machine precision.
Even if
$\tau'=\Delta t$, we obtain $x= 3.29\cdot 10^{-17}$.


\section{Initializing on a slow manifold} \label{INITI}
Remarkably, the modification we presented as \MdrieS above to improve the
approximation of coarse-scale \emph{steady state} solutions can form
the basis of an algorithm for appropriately initializing our
fine-scale simulators given a desired value of the observable(s). 
In the context of our simple example, this
means finding the point $(x^0,y)$ on the slow manifold corresponding to some prescribed value of $x$ we denote $x^0$.
We already showed how to use the constrained runs functional iteration (used in \Mtwee)
for this purpose.
%
The fixed point
of the constrained runs functional iteration lies close to (but, if
$x^0$ is not a coarse-scale steady state solution, not exactly on) the slow
manifold. 
More accurate  initializations can be obtained by using
variants of the constrained runs functional iteration outlined in the literature
(\cite{kreiss,lorenz,curry,aenio,ptasm,cmalc})
that  solve
$\textrm{d}^{m+1} y(x^0,y)/\textrm{d}t^{m+1} =0$ for a certain value of $m \in
\mathbb{N}$.
The larger the value of $m$, the more accurate the procedure will be
(\cite{ptasm}), assuming it converges.
For these variants of the scheme, however, the constrained runs
functional iteration is not guaranteed to converge to a solution close to the
slow manifold, and the computational complexity may be unacceptably large. 
For
these reasons, we now propose an alternative initialization method that has many similarities with
\Mdrie, but now, instead of computing coarse-scale steady state
solutions, we compute points lying on the slow manifold for a given value of the observable, $x^0$.

Instead of demanding that the finite difference approximation of the time derivative of the observable 
is zero after a simulation over time $\tau$ as in \eqref{FPF3}, we now compute $x$ so that
\begin{equation}
\label{init}
\Phi(x,\tau)-x^0=0.
\end{equation}
As in \Mdrie, $\Phi(x,\tau)$ denotes a coarse time-step over time $\tau$ in
which we use the simple arbitrary lifting scheme. 
Again, the value of $\tau$ should
be large enough to allow $y$ to get slaved to $x$, but
much smaller than the  coarse time scales.
If $\overline{x}$ is the solution of \eqref{init},
we expect
the fine-scale solution $(x,y)$, obtained after simulation over time $\tau$
starting from the arbitrary lifted state corresponding to $\overline{x}$, to be
a good approximation of the desired point on the slow manifold (the
simulation step has brought us close to the slow manifold at the desired value
$x^0$).
The obvious advantages of this algorithm (which we will refer to 
as \Minit) compared to the constrained runs functional iteration
are that this method is
conceptually  simpler, that it does not have the same potential
convergence issues, and that we are no
longer solving a system of equations of the form \mbox{$\textrm{d} y(x^0,y) / \textrm{d}t =0$}
in the space of the ``remaining variables'' (which may be
an advantage if that space is large compared to the space of the observables).
We still do, of course, require a solver for equation  \eqref{init}, as would also
be required for equation (\ref{FPF12}) above; this might
be something as simple as a Newton or Broyden method, or as sophisticated as
a Newton-Krylov GMRES solver (\cite{Kelley_Book}).



To illustrate the performance of the (variants of the) constrained runs
functional iteration and  \Minit,  we again consider the model problem
\eqref{stepper} with $\alpha=\pi/6$ and $\beta = -\pi/3$.  
Using five different values of $m$
for the constrained runs functional iteration, and also using \Minit, 
we approximate the value of $y$ so that $(1,y)$ lies as close
as possible to the exact point on the slow manifold, $(1,1/\sqrt{3})$.
Table \ref{Tab1} shows, for various values of $m$, the error in the solution found by
the constrained runs functional iteration
with  $\tol=10^{-16}$, and also the error found by \Minit (solved with a Newton iteration).
We clearly observe that, as the value of $m$ increases,  the error decreases
by a factor of about $(1-0.999)/(1-0.1)\approx 1.11 \cdot 10^{-3}$,
as expected by theory (\cite{ch_thesis}).
The errors in the solution found by both the $m=4$ constrained runs functional iteration
and \MinitS  with $\tau=15 \Delta t$ are at the
level of machine precision.

\begin{table} \caption{Absolute value of the error in the solution of the
constrained runs functional iteration and \Minit. \label{Tab1}}
 \begin{minipage}{0.49\textwidth}
\begin{center}
  \begin{tabular}{l |c}
 & $|$error$|$  \\
\hline
CRFI, $m=0$ & $8.55 \cdot 10^{-04}$ \\
CRFI, $m=1$ & $9.50 \cdot 10^{-07}$\\
CRFI, $m=2$ & $1.06 \cdot 10^{-09}$\\
\end{tabular}
\end{center}
 \end{minipage}
\begin{minipage}{0.49\textwidth}
\begin{center}
  \begin{tabular}{l |c}
 & $|$error$|$   \\
\hline
CRFI, $m=3$ & $1.17 \cdot 10^{-12}$\\
CRFI, $m=4$ & $3.11 \cdot 10^{-15}$\\
\Minit & $2.22 \cdot 10^{-16}$\\
\end{tabular}
\end{center}
 \end{minipage}

\end{table}

\section{Application to a lattice Boltzmann model}
\label{numerics}


In this section, we will apply the algorithms we have presented
to a lattice Boltzmann model
(LBM) of a one-dimensional reaction-diffusion system. 
In Section \ref{S:LBM},
we present the LBM. 
In Section \ref{anal} we analytically derive two
reduced models in terms of two different coarse observables. 
In Section
\ref{S:LBMres}, a detailed description of the numerical results is given.

\subsection{The lattice Boltzmann model}
\label{S:LBM}
An LBM (\cite{caalb}) describes the evolution of discrete (particle) distribution
functions  $f_i(x_j,t_k)$, which
depend on space $x_j$, time $t_k$ and velocity $v_i$. 
For our one-dimensional
model problem, only three values are considered for the velocity ($v_{i} = i
\Delta x / \Delta t$, with $i \in \{-1,0,1\}$),
and each distribution function $f_i$
is discretized in space on the domain $[0,1]$ using a grid spacing
$\Delta x=1/N$ ($N$ lattice intervals) and in time using a
time step $\Delta t$.
The LBM
evolution law for the distributions $f_i(x_j,t_k)$ in the interior of the
domain is
\begin{equation} \label{stdLBvgl1}
\begin{split}
f_i(x_{j+i},t_{k+1})
&=f_i(x_j + i \Delta x, t_k + \Delta t)
\\&= f_i(x_j,t_k) -
\omega  \left( f_i(x_j,t_k) - f_i^{eq}(x_j,t_k) \right) +
\frac{\Delta t}{3} F(\rho(x_j,t_k)),
\end{split}
\end{equation}
with $i \in \{-1,0,1\}$.

Diffusive collisions are modeled by the Bhatnagar-Gross-Krook (BGK) collision
term  
$- \omega (f_i(x_j,t_k) -
f_i^{eq}(x_j,t_k))$ 
as a relaxation to the local
diffusive equilibrium (\cite{sidra})
\begin{equation} \label{feq1}
f^{eq}_i(x_j,t_k) = \frac{1}{3} \rho(x_j,t_k).
\end{equation}
The parameter $\omega \in (0,2)$ is called the relaxation
coefficient and $\rho$ is the
(particle) density field,  which is defined as
the ``zeroth'' order velocity moment of $f_i(x_j,t_k)$
\begin{equation*} \label{rho}
\rho(x_j,t_k) = \sum_{i=-1}^1
f_i(x_j,t_k)=f_{-1}(x_j,t_k)+f_0(x_j,t_k)+f_1(x_j,t_k).
\end{equation*}
It follows directly that the BGK diffusive collisions locally
conserve
density.

The last term in equation \eqref{stdLBvgl1} models the reactions, which are
assumed   to depend only on the density field $\rho$ (\cite{sidra,lbcfr}). 
In
this paper, we will use the specific reaction term
\begin{equation}
\label{reactionterm}
 F(\rho(x_j,t_k))=\lambda \rho(x_j,t_k) \left( 1-\rho(x_j,t_k) \right),
\end{equation}
in which the parameter $\lambda
\geq 0$ determines the strength of the reaction ``force''.
Nonlinear reaction terms of the form \eqref{reactionterm} arise naturally in the
fields of heat and mass transfer (\cite{mmoha}) or in ecology (\cite{pdeie}).

At the boundaries, we impose Dirichlet boundary
conditions $\rho(0,t_k)=\rho(1,t_k)=0$
by assigning the appropriate values to the
 distribution functions that stream into the domain at $x_0=0$ and $x_N=1$.


Similar to the density $\rho$, we can  define the momentum $\phi$ and the
energy $\xi$ as (a rescaling of) the first and the second  (or in
short, the higher) order moments of $f_i$
\begin{equation*} \label{HOM}
\begin{split}
\phi(x_j,t_k) = \sum_{i=-1}^1 i f_i(x_j,t_k)=-f_{-1}(x_j,t_k) +f_{1}(x_j,t_k),
\\ \xi(x_j,t_k) =
\frac{1}{2}
\sum_{i=-1}^1 i^2 f_i(x_j,t_k)= \frac{f_{-1}(x_j,t_k)+f_1(x_j,t_k)}{2}.
\end{split}
\end{equation*}
Later on we will also use the variable $\sigma$, which is defined as
\begin{equation*} \label{sigma}
\sigma(x_j,t_k)  =\sum_{i=-1}^1 (2i^2-1)
f_i(x_j,t_k) = f_{-1}(x_j,t_k) -f_0(x_j,t_k)+ f_1(x_j,t_k) = -\rho(x_j,t_k)+4 \xi(x_j,t_k).
\end{equation*}


\subsection{Analytical coarse-graining} \label{anal}

For the LBM described in the previous section,
a Chapman-Enskog multiscale expansion can be used to
derive an accurate reduced model for the long-term behavior of the system
(\cite{caalb,tlbef}). 
(In practice, the equation-free methods should of course
only be used when such a derivation is not possible. Here, however, the
analytical derivation provides insight into the problem, which is particularly
helpful in understanding the performance of
the different methods.) 
In this section, we will derive two different
reduced models: one in terms of the density
$\rho$ and another in terms of the variable $\sigma$.
For a detailed derivation of the following two equations, we refer to appendix B of \cite{ch_thesis}.

If we use the density $\rho$ as the observable, the reduced model is  the
partial differential equation
\begin{equation}
\label{macroeq_rho}
\frac{\partial \rho(x,t)}{\partial t}=D \frac{\partial^2 \rho(x,t)}{\partial
x^2}+F(\rho) 
=\left ( \frac{2-\omega}{3\omega}
\frac{\Delta x^2}{\Delta t}  \right ) \frac{\partial^2 \rho(x,t)}{\partial
x^2}+ F(\rho) 
\end{equation}
with Dirichlet boundary conditions $\rho(0,t)=\rho(1,t)=0$.  
If we consider the
variable $\sigma$ to be the observable, we obtain the partial differential
equation
\begin{equation}
\label{macroeq_sigma}
\frac{\partial \sigma(x,t)}{\partial t}=D \frac{\partial^2
\sigma(x,t)}{\partial
x^2}+ \frac{1}{3} F(3\sigma) 
=\left ( \frac{2-\omega}{3\omega}
\frac{\Delta x^2}{\Delta t}  \right ) \frac{\partial^2 \sigma(x,t)}{\partial
x^2}+  \frac{1}{3} F(3\sigma) 
\end{equation}
with Dirichlet boundary conditions $\sigma(0,t)=\sigma(1,t)=0$.
%
%
%
%
%

As a by-product of the Chapman-Enskog expansion, we find, after dropping the
indices $j$ and $k$, and retaining only terms up to second order, that the
relation between $\sigma$ and $\rho$ is given by \begin{equation}
\label{third}\begin{split} \sigma(x,t) &= \frac{\rho(x,t)}{3} +
\frac{2(2-\omega)}{9\omega^2}\frac{\partial^2
\rho(x,t)}{\partial x^2} \Delta x^2+ \mathcal{O}(\Delta x^3) \\
\rho(x,t) &= 3 \sigma(x,t) -  \frac{2(2-\omega)}{\omega^2}\frac{\partial^2
\sigma(x,t)}{\partial x^2} \Delta x^2+ \mathcal{O}(\Delta x^3).
\end{split}
\end{equation}
%
This shows that in a LBM simulation,
the value of $\rho$ is approximately three times as large as the value of
$\sigma$, at least after a short initial transient. 
From
this point of view, the choice of $\sigma$ as the observable is as natural as
the choice of $\rho$.


The fact that the reduced dynamics can be described in terms of $\rho$ or
$\sigma$ only implies that the remaining ``fine-scale variables'' $\phi$ and $\xi$
quickly become functionals of (slaved to) $\rho$ or $\sigma$. 
These functionals,
which we will call slaving relations, are
 \begin{equation}
 \label{slaving_mom}
 \phi(x,t)=  -\frac{2}{3\omega} \frac{\partial \rho(x,t)}{\partial
 x} \Delta x + 
\mathcal{O}(\Delta x^3)
 \end{equation}
 \begin{equation}
  \label{slaving_ene}
 \xi(x,t) = \frac{1}{3}\rho(x,t)
 -\frac{\omega-2}{18
 \omega^2} \frac{\partial^2 \rho(x,t)}{\partial
 x^2} \Delta x^2 + \mathcal{O}(\Delta x^3)
 \end{equation}
in terms of the density $\rho$, or
\begin{equation}
 \label{slaving_mom_sigma}
 \phi(x,t)=  -\frac{2}{\omega} \frac{\partial \sigma(x,t)}{\partial
 x} \Delta x + \mathcal{O}(\Delta x^3)
 \end{equation}
 \begin{equation}
  \label{slaving_ene_sigma}
 \xi(x,t) = \sigma(x,t)
 +\frac{\omega-2}{2 \omega^2} \frac{\partial^2 \sigma(x,t)}{\partial
 x^2} \Delta x^2 + \mathcal{O}(\Delta x^3)
 \end{equation}
in terms of the variable $\sigma$.
 The slaving relations \eqref{slaving_mom}-\eqref{slaving_ene} or
\eqref{slaving_mom_sigma}-\eqref{slaving_ene_sigma}
 define a slow manifold in the LBM phase space, on which the reduced dynamics
takes place.

\begin{figure}
\begin{center} \includegraphics[width=\textwidth]{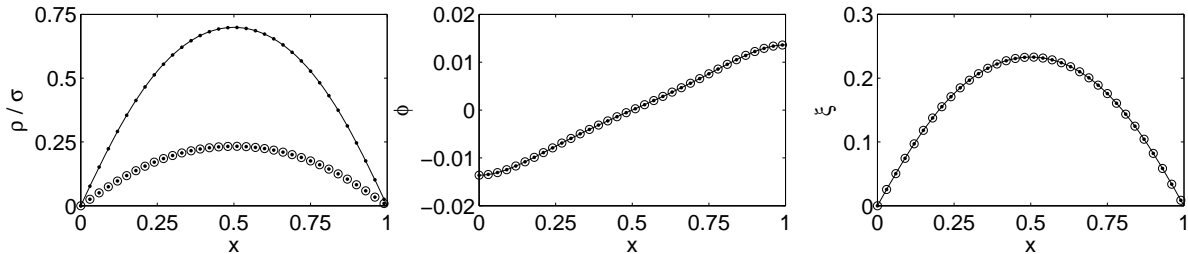}
\caption{The full LBM steady state solution, computed in both the $(\rho,\phi,\xi)$- and the $(\sigma,\phi,\xi)$-coordinate systems using solid line and circle markers, respectively.  This solution is shown in $\rho/\sigma$ space (left), $\phi$ space (middle), and $\xi$ space (right).  Dot markers are also present to demonstrate the validity of \eqref{macroeq_rho} and \eqref{macroeq_sigma} (left), \eqref{slaving_mom} and \eqref{slaving_mom_sigma} (middle), and \eqref{slaving_ene} and \eqref{slaving_ene_sigma} (right). \label{Fig2}}
\end{center}
\end{figure}

The validity  of the reduced models
\eqref{macroeq_rho}-\eqref{macroeq_sigma} and the slaving relations \eqref{third}--\eqref{slaving_ene_sigma} is illustrated in Figure \ref{Fig2}. 
Here, we chose $N=100$, $\lambda=25$ and $\omega=1.25$, and computed the full LBM steady state
solution in both the $(\rho,\phi,\xi)$- and the $(\sigma,\phi,\xi)$-coordinate systems.
These solutions are shown using the solid line and the circle markers,
respectively. 
To simplify the interpretation, we split the solution into the $\rho$-, $\sigma$-, $\phi$- and $\xi$-components,
and mapped them onto the spatial domain $[0,1]$.
From these solutions, it is confirmed that the value of $\sigma$ is indeed about one third of the value of
$\rho$.
Then, we also computed the steady state solutions of
\eqref{macroeq_rho} and \eqref{macroeq_sigma}, and added them in the form of the dot markers to Figure
\ref{Fig2} (left). 
At the resolution shown here, these reduced solutions are clearly
indistinguishable from the $\rho$- and $\sigma$-components of the full LBM
solution.
Finally, we added to Figure  \ref{Fig2} (middle and right), also in the form of the dot markers, the profiles of $\phi$ and $\xi$ according to the slaving relations \eqref{slaving_mom}-\eqref{slaving_ene} (exactly the same results are obtained when using \eqref{slaving_mom_sigma}-\eqref{slaving_ene_sigma}). 
Here,
we  approximated the spatial derivatives of $\rho$ numerically using
finite differences. 
In this case also, the values of $\phi$ and $\xi$ are indistinguishable from
the $\phi$- and $\xi$-components of the full LBM solution.

Any LBM initial condition is quickly attracted towards the slow manifold during
the transient LBM simulation. 
For $\omega=1.25$, for instance, the slow manifold is reached, up to machine precision, 
in about 25 LBM time steps (the values of $\phi$ and $\xi$ in the simulation converge linearly to the
slaving relations with convergence factor $|1-\omega|$, as can be seen from \eqref{stdLBvgl1}).
The value of the observable may however change during the
 transient phase towards the slow manifold:
%
%
\begin{itemize}
 \item Suppose that we start from the initial condition
$(\rho,\phi,\xi)=(\rho_0,0,0)$. 
In terms of the distribution functions,
this corresponds to $(f_{-1}, f_0, f_1) = (0, \rho_0, 0)$. 
During the initial
 transient phase, the values of the distribution functions will quickly be
redistributed to $(f_{-1}, f_0, f_1) \approx (\rho_0/3, \rho_0/3, \rho_0/3)$,
or, in terms of the velocity moments, to
$(\rho,\phi,\xi)\approx(\rho_0,0,\rho_0/3)$ (this follows directly from
\eqref{stdLBvgl1}-\eqref{feq1} and fact that the evolution towards the slow
manifold is very fast compared to the diffusive or reactive time scales). 
Hence,
the density $\rho$ \emph{does not change substantially} during the transient phase.
\item 
Suppose that we start from the initial condition
$(\sigma,\phi,\xi)=(\sigma_0,0,0)$. 
In terms of the distribution functions,
 this corresponds to $(f_{-1}, f_0, f_1) = (0, -\sigma_0, 0)$. 
During the
initial
 transient phase, the values of the distribution functions will quickly be
redistributed to $(f_{-1}, f_0, f_1) \approx (-\sigma_0/3, -\sigma_0/3,
-\sigma_0/3)$,
or, in terms of $\sigma$ and the higher order velocity moments, to
$(\sigma,\phi,\xi)\approx(-\sigma_0/3,0,-\sigma_0/3)$. 
Hence,
the variable $\sigma$ \emph{does change substantially} during the transient phase from
 $\sigma_0$ to $-\sigma_0/3$.
\end{itemize}
As we demonstrated before (and will see below), the fact that the observable changes substantially during
the transient phase towards the slow manifold may have important
consequences for the performance of the different methods.

\subsection{Numerical results}
\label{S:LBMres}
In all numerical experiments reported below, we use $D=1$, $\Delta x=1/N=1/100$
and
$\omega=1.25$. 
According to \eqref{macroeq_rho}-\eqref{macroeq_sigma}, the value
of the LBM time step is then $\Delta t=2 \cdot 10^{-5}$. 
For $\lambda$, we
will choose  $\lambda=25$ or $\lambda=5$.
For $\lambda=25$, the LBM \eqref{stdLBvgl1}--\eqref{reactionterm} exhibits a
nontrivial \emph{stable} steady state solution; for $\lambda=10$, the nontrivial
steady state solution is \emph{unstable} (\cite{ch_thesis}).
These solutions will henceforth
be referred to as ``the stable steady state solution" and ``the unstable steady state solution", respectively.

To solve the nonlinear systems \eqref{FPF12}, \eqref{FPF3} or \eqref{init},
which we will write below in more abstract form as $g(x)=0$, we use the most basic implementation of Newton's method
in which we estimate the required  Jacobian matrices $\partial g / \partial x
(x_i)$
in a column-by-column fashion.
Note that these Jacobian matrices  are typically
not very large, as their size is determined by the dimension of the
\emph{coarse} subspace; still, the cost of the computational linear algebra should be taken into consideration.
Column $l$ of $\partial g / \partial x (x_i)$ is the directional
derivative in the direction of the $l$-th unit vector $e_l$, which can be
approximated as
\begin{equation}
 \label{numest} \frac{\partial g}{\partial x} (x_i)\cdot e_l
\approx
 \frac{g(x_i+\varepsilon e_l) -g(x_i)}{\varepsilon},
\end{equation}
with $\varepsilon$ an appropriate small parameter. 
The
resulting linear systems $\partial g / \partial x (x_i) \Delta x = -g(x_i)$ are then solved by Gaussian elimination. 
For problems for which the coarse
subspace is very large, it may be more appropriate to solve the linear
systems using, for instance, a Jacobian-free Newton-Krylov method (\cite{Kelley_Book}).
The use of
these advanced methods will not be considered in this article.

%

\subsubsection{Results using $\rho$ as the observable}
In this section, we apply our algorithms and the constrained runs
functional iteration to the LBM, using the density $\rho$ as the
observable. 
As $\rho$ does not change substantially during the
transient phase towards the slow manifold, we expect to obtain good results
with all methods.
Note that the results reported here are in perfect correspondence with the theoretical results obtained in \cite{ch_25}, in which it was shown that the coarse time-stepper under consideration is actually a time integrator for a (slightly) modified reaction-diffusion system.

\paragraph{Coarse-scale steady states: application of \Meen}
We approximate the $\rho$-component of the stable or unstable
 steady state solution with
\Meen, in which we use the arbitrary lifting scheme $\rho \mapsto (\rho,0,0)$
and set $\tau=25 \Delta t$. 
We choose $\varepsilon=\sqrt{\eta}$, with $\eta
\approx 2.22 \cdot 10^{-16}$ the machine precision. As the initial condition
for Newton's method, we take $\rho^0=\sin(\pi x)$ when $\lambda=25$ and
$\rho^0=-\sin(\pi x)$ when $\lambda=5$. 
For both values of $\lambda$, the
nonlinear residual reaches the level of machine precision after 5 Newton
steps; the resulting solution is called $\rho_1$. 
Table \ref{tabel1} shows the
two-norm of the error $||e||_2=||\rho_1-\rho^*||_2$ for $\lambda=25$  and
$\lambda=5$ (as before, we use $\rho^*$ to denote the $\rho$-component of the full LBM
steady state solution). 
The error is not large, but it is also not small (the relative
error is about 2\% when $\lambda=25$ and about 6\% when $\lambda=5$).  
This can be explained
by the fact that $\rho$ \emph{does change}, although very slightly, during the fast
transient phase.
Note that changing $\tau$ will also influence the approximation of the steady-state solution.

\begin{table}
\caption{Two-norm of the error
$||e||_2=||\rho_i-\rho^*||_2$ in the steady state solution for \MeenS ($i=1$),
\MtweeS ($i=2$), and \MdrieS ($i=3$), using $\lambda=25$ and $\lambda=5$.  $\rho^*$ is known
analytically for each $\lambda$.
\label{tabel1}}
\begin{center}
\begin{tabular}{l |ccc}
 & \Meen & \Mtwee & \Mdrie \\
\hline
$\lambda=25$ (stable) & $1.01 \cdot 10^{-1}$  & $ 1.07 \cdot 10^{-13}$
& $2.58\cdot 10^{-13}$ \\
$\lambda=5$ (unstable) &  $4.79 \cdot 10^{-1}$ & $1.47\cdot 10^{-12}$
&$1.85\cdot 10^{-12}$  \\
\end{tabular}
\end{center}
\end{table}

\paragraph{Coarse-scale steady states: application of \Mtwee}
We now approximate the $\rho$-component of the stable or unstable
 steady state solution with
\Mtwee, in which we set $\tau= \Delta t$ and use the constrained runs lifting
scheme with $m=0$ and $\tol=10^{-14}$, starting from the initial condition
$(\phi,\xi)=(0,0)$.  
Again, we choose $\varepsilon=\sqrt{\eta}$, and use
 $\rho^0=\sin(\pi x)$   or
$\rho^0=-\sin(\pi x)$ as the initial condition for Newton's method.
In \cite{maote}, it was shown that for $m=0$ and  $\rho$ as the observable,
the eigenvalues of the constrained runs iteration matrix lie on a circle with
center point 0 and radius $|1-\omega|$. 
As a consequence, the iteration is
stable for all values of $\omega \in (0,2)$ and for $\omega=1.25$ the
convergence factor is 0.25 (in other words, about 25 iterations are needed to
reach the tolerance $\tol$).
%
  As in \Meen,  the nonlinear residual reaches the level of
machine precision after 5 Newton steps for both values of $\lambda$; the
resulting solution is now called $\rho_2$. 
Table \ref{tabel1} shows the two-norm
of the error $||e||_2=||\rho_2-\rho^*||_2$  for  $\lambda=25$  and
$\lambda=5$. 
The error is now clearly very small.


\paragraph{Coarse-scale steady states: application of \Mdrie}

We approximate the $\rho$-component of the stable or unstable steady state
solution with \Mdrie, in which we  use the arbitrary lifting scheme $\rho
\mapsto (\rho,0,0)$ and set $\tau=25 \Delta t$ and $\tau'=\Delta t$.
To avoid numerical complications due to (nearly) singular Jacobian matrices, we
explicitly eliminate the boundary conditions $\rho(0)=\rho(1)=0$ (so the
dimension of the Jacobian matrix $\partial g / \partial x (x_i)$ is ($N-1$)$\times$($N-1$) instead of
($N+1$)$\times$($N+1$)).
We also set $\varepsilon=\sqrt[4]{\eta}$ (see also the
discussion in Section \ref{err_numerics}).
%
As the initial condition for Newton's method, we again use $\rho^0=\sin(\pi
x)$ or $\rho=-\sin(\pi x)$.  
For both values of $\lambda$, the nonlinear
residual reaches the level of machine precision after about 6 or 7 Newton
steps; the resulting solution is now called  $\overline{\rho}_3$. 
Table
\ref{tabel1} shows the two-norm of the error $||e||_2=||\rho_3-\rho^*||_2=
||\Phi(\overline{\rho}_3, \tau)-\rho^*||_2$  for $\lambda=25$  and
$\lambda=5$. 
Again, the error is very small.

\begin{table} \caption{Two-norm of the error in the solution of the
constrained runs functional iteration and \Minit, when $\rho$ is the
observable. \label{tabel2}} \begin{center} \begin{tabular}{l |c  c}
 & $\lambda=25$  &  $\lambda=5$ \\
\hline
CRFI, $m=0$ & $ 9.97 \cdot 10^{-5}$ & $ 7.00 \cdot 10^{-5}$\\
CRFI, $m=1$ & $ 2.47\cdot 10^{-7}$&$  2.59\cdot 10^{-7}$\\
CRFI, $m\geq 2$ & $\infty$& $\infty$\\
\Minit & $ 1.03\cdot 10^{-15}$&$1.19 \cdot 10^{-15}$\\
\end{tabular}
\end{center}
\end{table}

\paragraph{Initialization on the slow manifold: the constrained runs functional iteration and \Minit}

To test how the constrained runs functional iteration and \MinitS 
perform when we attempt to initialize on the slow manifold,
we set
up the following experiment. 
For both $\lambda=25$ and $\lambda=5$, we first
 perform a LBM simulation of 50 steps, starting from
 the (arbitrary) initial condition $(\rho^0,\phi^0,\xi^0)=(x(1-x),x,\sin(\pi x))$.
 This provides us with a
LBM state $(\rho^*,\phi^*,\xi^*)$ ``on" the slow manifold which is not a coarse
scale steady state. 
Then, we use the
constrained runs functional iteration 
or \MinitS to approximate the values of $\phi^*$ and $\xi^*$ corresponding to
$\rho^*$.
For the constrained runs functional iteration, we use various values of $m$,
set $\tol=10^{-14}$ and start from the initial condition $(\phi,\xi)=(0,0)$.
For \Minit, we set $\tau=25 \Delta t$, start from the initial condition
$\rho^0=\rho^*$, and again explicitly eliminate the boundary
conditions $\rho(0)=\rho(1)=0$ to avoid
numerical complications due to (nearly) singular Jacobian matrices.
We also set $\varepsilon=\sqrt[4]{\eta}$ (see also the
discussion in Section \ref{err_numerics}).

The results are summarized in Table \ref{tabel2}. 
For both the constrained runs
functional iteration and \MinitS, and for $\lambda=25$ and
$\lambda=5$, we tabulate the two-norm of the error
$||e||_2=||(\phi^{\#},\xi^{\#}) -(\phi^*,\xi^*)||_2$ (we use
$(\phi^{\#},\xi^{\#})$ to denote the solution found by the constrained runs
functional iteration or \Minit).
The solution of the constrained runs functional
iteration with $m=1$ is more accurate than the solution obtained when $m=0$,
but for values of $m\geq2$ the iteration is {\it unstable} (some of the
eigenvalues of the constrained runs iteration matrix are larger than 1 in
magnitude).
For \Minit, the nonlinear residual reaches the level of machine precision
after about 3 or 4 Newton steps; the resulting solution is called
$\overline{\rho}_4$. 
In this case, $\phi^{\#}$ and $\xi^{\#}$ are the
values of the higher order moments $\phi$ and $\xi$ obtained after a LBM
simulation over time $\tau$ starting from $(\overline{\rho}_4,0,0)$.
The solution found by \MinitS is clearly very accurate.


%


\subsubsection{Results using $\sigma$ as the observable}
In this section, we apply our algorithms as well as the constrained runs functional iteration
to the LBM, this time using the variable $\sigma$ as the
observable. 
As $\sigma$ \emph{does change} substantially during
the transient phase towards the slow manifold, we expect to obtain poor
results with \Meen, \Mtwee, and the constrained runs functional iteration, but
good results with \MdrieS and \Minit.
Tables \ref{tabel3} and \ref{tabel4} are the analogues of Tables \ref{tabel1}
and \ref{tabel2}; they tabulate the results of the numerical experiments described below.
Note that the results reported here are in perfect correspondence with the theoretical results obtained in \cite{ch_25}, in which it was shown that the coarse time-stepper under consideration is actually a time integrator for a (slightly) modified reaction-diffusion system.


\paragraph{Coarse-scale steady states: application of \Meen}
We approximate the $\sigma$-component of the stable or unstable
 steady state solution with
\Meen, in which we use the arbitrary lifting scheme $\sigma \mapsto
(\sigma,0,0)$ and set $\tau=25 \Delta t$. 
As before, we choose
$\varepsilon=\sqrt{\eta}$. 
As the initial condition for Newton's method, we
use $\sigma^0=\sin(\pi x)/3$ when $\lambda=25$ or $\sigma^0=-\sin(\pi x)/3$
when $\lambda=5$ (this choice is motivated by the fact that on the slow
manifold, the value of $\sigma$ is about one third of the value of $\rho$;
cf.\ \eqref{third}).
For both values of $\lambda$, the nonlinear residual reaches the level of
machine precision after 3 Newton steps; the resulting solution is called
$\sigma_1$. 
Table \ref{tabel3} shows the two-norm of the error
$||e||_2=||\sigma_1-\sigma^*||_2$ for $\lambda=25$  and $\lambda=5$ (as before, we use
$\sigma^*$ to denote the $\sigma$-component of the full LBM steady state
solution). 
The error is clearly unacceptable (the iteration converges to $\sigma=0$).

\begin{table}
\caption{
Two-norm of the error
$||e||_2=||\sigma_i-\sigma^*||_2$ in the steady state solution for \Meen ($i=1$),
\MtweeS ($i=2$) and \Mdrie ($i=3$), and using $\lambda=25$ and $\lambda=5$.  $\sigma^*$ is known
analytically for each $\lambda$.
\label{tabel3}}
\begin{center}
\begin{tabular}{l |ccc}
 & \Meen & \Mtwee & \Mdrie \\
\hline
$\lambda=25$ (stable) & $ 1.70$  & $ \infty$
& $3.77\cdot 10^{-13}$ \\
$\lambda=5$ (unstable) &  $ 2.69$ & $\infty$
&$6.16\cdot 10^{-13}$  \\
\end{tabular}
\end{center}
\end{table}

\paragraph{Coarse-scale steady states: application of \Mtwee}
We approximate the $\sigma$-component of the stable or unstable
 steady state solution with
\Mtwee, in which we set $\tau= \Delta t$ and use the constrained runs lifting
scheme with $m=0$ and $\tol=10^{-14}$, starting from the initial condition
$(\phi,\xi)=(0,0)$. 
Again, we choose $\varepsilon=\sqrt{\eta}$, and use
 $\sigma^0=\sin(\pi x)/3$   or
$\sigma^0=-\sin(\pi x)/3$ as the initial condition for Newton's method.
%
%
%
For $m=0$, $\omega=1.25$ and $\lambda=0$ (we use $\lambda=0$ rather than
$\lambda=25$ or $\lambda=5$ as the iteration is then linear), and using
$\sigma$ as the observable,  the eigenvalues of the constrained runs iteration
matrix lie in 
  $(-1.416,-0.25) \cup (0.25,1.416)$.
Also for nonzero values of $\lambda$, the eigenvalues of the (varying)
iteration matrix fall outside the unit circle. 
As a consequence, the iteration
is {\it unstable} and  \MtweeS cannot be used; it cannot even be started  
(see Table \ref{tabel3}).

\paragraph{Coarse-scale steady states: application of \Mdrie}

We approximate the $\sigma$-component of the stable or unstable steady state
solution with \Mdrie, in which we  use the arbitrary lifting scheme $\sigma
\mapsto (\sigma,0,0)$ and set $\tau=25 \Delta t$ and $\tau'=\Delta t$.
%
%
To avoid numerical complications due to (nearly) singular Jacobian matrices,
we again explicitly eliminate the boundary conditions $\sigma(0)=\sigma(1)=0$
and set $\varepsilon=\sqrt[4]{\eta}$ (see also the
discussion in Section \ref{err_numerics}).
%
%
%
%
%
As the initial condition for Newton's method, we now use $\sigma^0=-\sin(\pi
x)$ or $\sigma^0=\sin(\pi x)$  (this choice is motivated by the fact that,
after the initial transient towards the slow manifold, the value of $\sigma$
is about minus one third of the value of $\sigma^0$, so that we then end up
near $\sigma=\sin(\pi x)/3$ and $\sigma=-\sin(\pi x)/3$; cf.\ the last
paragraph in Section \ref{anal}).
 For both values of $\lambda$, the nonlinear
residual reaches the level of machine precision after about 8 ($\lambda=25$) or 20 ($\lambda=5$)
Newton iteration steps; the resulting solution is
now called  $\overline{\sigma}_3$. 
Table \ref{tabel3} shows the two-norm of
the error $||e||_2=||\sigma_3-\sigma^*||_2= ||\Phi(\overline{\sigma}_3,
\tau)-\sigma^*||_2$  for $\lambda=25$  and $\lambda=5$. 
Again, the error is
very small.

\paragraph{Initialization on the slow manifold: the constrained runs functional iteration and \Minit}

We now turn again to the problem of initializing on the slow manifold.
To test the constrained runs functional iteration and \Minit, we set up the
following experiment. 
For both $\lambda=25$ and $\lambda=5$, we first
 perform a LBM simulation of 50 steps, starting from
 the (arbitrary) initial condition $(\sigma^0,\phi^0,\xi^0)=(x(1-x),x,\sin(\pi x))$.
 %
 This provides us with a LBM state $(\sigma^*,\phi^*,\xi^*)$
``on" the slow manifold. 
Then, we use the
constrained runs functional iteration 
or \MinitS to approximate the values $\phi^*$ and $\xi^*$ corresponding to
$\sigma^*$.
For the constrained runs functional iteration, we use various values of $m$,
set $\tol=10^{-14}$ and start from the initial condition $(\phi,\xi)=(0,0)$.
For \Minit, we set $\tau=25 \Delta t$, start from the initial condition
$\sigma^0=-3\sigma^*$ (this choice is again motivated by the fact that, after
the initial transient towards the slow manifold, the value of $\sigma$ is
about minus one third of the value of $\sigma^0$, so that we then end up near
$\sigma=\sigma^*$; cf.\ the last paragraph in Section \ref{anal}) and again
explicitly eliminate the boundary conditions $\sigma(0)=\sigma(1)=0$ and set
$\varepsilon=\sqrt[4]{\eta}$ to avoid numerical complications due to (nearly)
singular Jacobian matrices (see also the
discussion in Section \ref{err_numerics}).

The results are summarized in Table \ref{tabel4}. 
For both the constrained
runs functional iteration and \Minit, and for $\lambda=25$ and $\lambda=5$, we
tabulate the two-norm of the error $||e||_2=||(\phi^{\#},\xi^{\#})
-(\phi^*,\xi^*)||_2$ (as before, we use $(\phi^{\#},\xi^{\#})$ to denote the solution
found by the constrained runs functional iteration or \Minit). 
As already
indicated above, the constrained runs functional iteration  is always {\it unstable}.
For \Minit, the nonlinear residual reaches the level of machine precision
after about 5 or 6 Newton steps; the resulting solution is called
$\overline{\sigma}_4$. 
In this case, $\phi^{\#}$ and $\xi^{\#}$ are the
values of the higher order moments $\phi$ and $\xi$ obtained after a LBM
simulation over time $\tau$ starting from
$(\overline{\sigma}_4,0,0)$. 
As before, the solution found by \MinitS
is extremely accurate (one cannot expect to do better, in fact).

\begin{table} \caption{Two-norm of the error in the solution of the
constrained runs functional iteration and \Minit, when $\sigma$ is the
observable. \label{tabel4}} \begin{center} \begin{tabular}{l |c  c}
 & $\lambda=25$  &  $\lambda=5$ \\
\hline
CRFI, $m=0$ & $\infty$ & $\infty$\\
CRFI, $m=1$ & $\infty$&$\infty $\\
CRFI, $m\geq 2$ & $\infty$& $\infty$\\
\Minit & $ 4.51\cdot 10^{-14}$&$1.12 \cdot 10^{-13}$\\
\end{tabular}
\end{center}
\end{table}

%

\section{Numerics} \label{err_numerics}


In all of the experiments above, we obtained accurate results with
\MdrieS and \Minit.  
As we explained, these methods do not suffer from inaccuracies introduced
by arbitrary lifting or from instabilities which sometimes accompany
the constrained runs functional iteration.
In some cases, however, numerical
difficulties may also be encountered when applying these methods. 
Let us illustrate this for \MdrieS using
$\rho$ as the observable. 
As before, we choose  $D=1$, $\Delta x=1/N=1/100$, $\omega=1.25$
and $\lambda=25$, use the arbitrary lifting scheme $\rho
\mapsto (\rho,0,0)$, set  $\tau'=\Delta t$, use $\rho^0=\sin(\pi
x)$ as the initial condition for Newton's method, and explicitly eliminate the boundary conditions $\rho(0)=\rho(1)=0$.
The value of $\varepsilon$, however, will now be set equal to
$\varepsilon=\sqrt{\eta}$ (with $\eta$ the value of machine precision)
instead of to $\varepsilon=\sqrt[4]{\eta}$, and we will vary $\tau$ from $0$ to $25 \Delta t$.

The results are summarized in Table \ref{tabel5}. 
As expected,
the  error (compared to the exact solution $\rho^*$)  decreases as the value of $\tau/\Delta t$ increases because larger $\tau$ values
bring the fine-scale simulator closer to the slow manifold.
However, except for very small values of $\tau/\Delta t$,
the condition numbers of the Jacobian matrices encountered within Newton's  method, $\kappa_i=|| \partial g / \partial x (x_i) ||\cdot || (\partial g / \partial x (x_i))^{-1}||$,  also tend to increase. 
This can be understood by realizing that some components of the observable $\rho$
are also decaying (at a fast pace, albeit slower than that of the remaining fine-scale variables), so that more and more relative indeterminacy is introduced  as the value of $\tau/\Delta t$ increases.
In other words, if we perturb the exact solution in the direction of a relatively quickly decaying coarse observable, the nonlinear residual will remain small
as this observable will largely be damped out by the time we reach the slow manifold.
Since the norm of the Jacobian matrix itself remains nearly constant, it is the norm of the
inverse of the Jacobian matrix that increases along with the condition number.

\begin{table} \caption{As a function of $\tau/\Delta t$, this table gives the number of Newton iteration steps required to
reach the tolerance $\tol=10^{-14}$,
the two-norm of the nonlinear residual $||\Phi(\overline{\rho}_3,\tau+\tau')-\Phi(\overline{\rho}_3,\tau)||_2$, the two-norm
of the error $||\Phi(\overline{\rho}_3,\tau)-\rho^*||_2$, the maximal value of the condition
number $\kappa_i$, and the maximal norm of the inverse of the Jacobian $|| (\partial g / \partial x (x_i))^{-1}||$
 encountered during the Newton iteration.\label{tabel5}} \begin{center} \begin{tabular}{c |c  c c c  c}
$\tau/\Delta t$ & \# iters.  &  residual & error &  $\max_i \kappa_i$ & $\max_i || (\partial g / \partial x (x_i))^{-1}||$\\
\hline
1 & 4  & 1.44e-015 & 1.73e+000 & 2.62e+003 & 3.77e+003 \\
2 & 6  & 7.05e-016 & 5.89e-001 & 5.25e+002 & 3.77e+003 \\
3 & 5  & 7.16e-016 & 1.94e-001 & 7.34e+002 & 3.24e+003 \\
4 & 5  & 1.03e-015 & 5.95e-002 & 2.37e+002 & 3.39e+003 \\
5 & 5  & 1.59e-015 & 1.77e-002 & 1.98e+002 & 3.38e+003 \\
6 & 5  & 9.36e-016 & 5.12e-003 & 1.70e+002 & 3.36e+003 \\
7 & 5  & 1.41e-015 & 1.45e-003 & 4.73e+002 & 1.07e+004 \\
8 & 5  & 1.30e-015 & 4.07e-004 & 1.42e+003 & 3.59e+004 \\
9 & 5  & 1.37e-015 & 1.12e-004 & 1.53e+002 & 4.26e+003 \\
10 & 5  & 5.98e-015 & 3.08e-005 & 8.18e+003 & 2.51e+005 \\
11 & 6  & 6.64e-016 & 8.37e-006 & 1.29e+004 & 4.29e+005 \\
12 & 5  & 5.25e-015 & 2.26e-006 & 1.88e+003 & 6.78e+004 \\
13 & 6  & 2.75e-015 & 6.06e-007 & 2.10e+005 & 8.15e+006 \\
14 & 6  & 1.18e-015 & 1.62e-007 & 1.33e+005 & 5.53e+006 \\
15 & 6  & 1.16e-015 & 4.30e-008 & 2.47e+004 & 1.09e+006 \\
16 & 8  & 2.19e-015 & 1.14e-008 & 5.15e+005 & 2.41e+007 \\
17 & 10  & 4.24e-015 & 3.01e-009 & 1.85e+006 & 9.21e+007 \\
18 & 10  & 2.10e-015 & 7.89e-010 & 3.82e+005 & 2.00e+007 \\
19 & 9  & NaN & NaN & 1.51e+008 & 4.12e+009 \\
20 & 3  & NaN & NaN & 6.12e+008 & 3.54e+010 \\
21 & 6  & NaN & NaN & 1.19e+008 & 7.20e+009 \\
22 & 6  & NaN & NaN & 1.22e+009 & 7.69e+010 \\
23 & 5  & NaN & NaN & 8.19e+009 & 3.80e+010 \\
24 & 3  & NaN & NaN & 2.11e+008 & 1.45e+010 \\
25 & 3  & NaN & NaN & 2.00e+009 & 1.43e+011 \\

\end{tabular}
\end{center}
\end{table}

As soon as the condition number and norm of the inverse of the Jacobian reach values larger than
$10^8$, we observe that Newton's method no longer converges.  
This can be explained as follows.
If $\Phi(x,\tau)$ and $\Phi(x,\tau+\tau')$ are both of $\mathcal{O}(1)$, the absolute error
in $g(x)$ above (remember, we are solving $g(x)=0$) is of $\mathcal{O}(\eta)$.
Due to round-off and truncation error in the finite difference approximation of the Jacobian, the absolute error in the elements of the Jacobian matrices $\partial g / \partial x(x_i)$
is then of  $\mathcal{O}(\eta/\varepsilon+\varepsilon)=\mathcal{O}(10^{-8})$.
To see the influence of this Jacobian matrix perturbation, we
can write
\begin{equation} \label{eqcond}
(\partial g / \partial x (x_i) + \delta g) (\Delta x+\delta x) =-g(x_i),
\end{equation}
in which $\delta g$ represents the perturbation matrix (with elements of 
$\mathcal{O}(10^{-8})$) and $\delta x$ represents the resulting perturbation on $\Delta x$. 
If we neglect the
 $\delta g \delta x$ term, we find that
\begin{equation} \label{eqcond2}
\delta x \approx -(\partial g / \partial x (x_i))^{-1} \cdot  \delta g \cdot \Delta
x.
\end{equation}
It is well known that our (inexact) Newton method converges if
\begin{equation}
\frac{|| \partial g / \partial x (x_i) (\Delta x+\delta x) +g(x_i) ||}{|| g(x_i)  ||} < 1,
\end{equation}
at least if we start sufficiently close to the solution (\cite{inexact}). 
Using
\eqref{eqcond2}, this becomes
\begin{equation}
\frac{|| \partial g / \partial x (x_i) (\Delta x- (\partial g / \partial x (x_i))^{-1} \cdot  \delta g \cdot \Delta
x ) +g(x_i) ||}{|| g(x_i)  ||} =  \frac{|| \delta g \cdot \Delta
x  ||}{|| g(x_i)  ||} \lesssim 1.
\end{equation}
If $||\cdot||$ denotes the Euclidean vector norm or its induced matrix norm (i.e., the spectral
norm), it holds that
\begin{eqnarray}
  \frac{|| \delta g \cdot \Delta
x  ||}{|| g(x_i) ||} &=& C \cdot \frac{|| \delta g || \cdot || \Delta
x  ||}{|| g(x_i) ||}  \\ &=& C \cdot D \cdot || \delta g || \cdot || (\partial g / \partial x (x_i))^{-1}
|| \notag \\ &=&  C \cdot D \cdot  \kappa_i \cdot \frac{|| \delta g ||}{|| \partial g / \partial x (x_i) ||}, \notag
\end{eqnarray}
with $C, D \in [0,1]$. 
(Here, we also used the fact that $ \Delta x = -(\partial g / \partial x (x_i))^{-1} \cdot g(x_i) \Rightarrow ||\Delta x|| = D \cdot ||(\partial g / \partial x (x_i))^{-1}|| \cdot
||g(x_i)||$.) Since the values $C$ and $D$ are in practice often approximately equal to $1$, it follows that Newton's method  converges
if
 \begin{equation}
   || \delta g || \cdot || (\partial g / \partial x (x_i))^{-1}
|| =   \kappa \frac{|| \delta g ||}{|| \partial g / \partial x (x_i) ||} \lesssim 1,
\end{equation}
at least if we start sufficiently close to the solution.
As $|| \delta g||=\mathcal{O}(10^{-8})$, this implies that  Newton's method is expected to  converge
%
%
if $|| (\partial g / \partial x (x_i))^{-1}|| < \mathcal{O}(10^8)$.
If $|| (\partial g / \partial x (x_i))^{-1}|| > \mathcal{O}(10^8)$, 
the method is expected to diverge.
These theoretical results are clearly confirmed in  Table \ref{tabel5}.


Note that for our LBM model problem, the error in the solution can be made as small as $\mathcal{O}(10^{-10})$
by using the largest value of $\tau/\Delta t$ for which Newton's method converges.
If the  desired level of accuracy cannot be reached due to numerical difficulties,
 one may try the following ``trick'':
 increase the value of $\varepsilon$, as we did in Section \ref{S:LBMres} when we set $\varepsilon = \sqrt[4]{\eta}$ instead of
$\varepsilon = \sqrt{\eta}$,
 in an attempt to reduce the error of the finite difference
 approximation. 
 Remember that the finite difference error consists of round-off error and truncation error so that the total
 finite difference error is of  $\mathcal{O}(\eta/\varepsilon+\varepsilon)$.
 If the problem is only mildly  nonlinear, as in the case of our LBM due to the fact that $\Delta t$ is small, the constant in the $\mathcal{O}(\varepsilon)$ term is small and the  finite difference error can be decreased by choosing $\varepsilon>\sqrt{\eta}$.


\section{Conclusions}

In this paper, we have introduced an approach for the computation of coarse-scale steady state solutions 
as well as an approach for initialization on a slow manifold.  
These methods were compared favorably to previously suggested ones: \MdrieS and \MinitS are quick,
accurate, and robust, and they bear a striking similarity to each other.  
We demonstrated the use of each of these methods
on a lattice Boltzmann model for a reaction-diffusion system, compared them to previously suggested methods, and verified our error predictions with both numerical results
and numerical analysis.  
These new procedures circumvent the need for long, fine-scale simulations to find coarse-scale
steady states, or to appropriately initialize the fine-scale simulator.

 Our implementation of the numerical methods in this report has been simple and direct, in order to clearly illustrate the methods and analyze their sources of error.  
 Indeed, in real-world applications, components like the use of higher order derivatives, the reusing of data (for example, for two nearby sets of observables, the corresponding fine-scale initializations are probably similar), or more intelligent Newton steps may clearly be used to improve performance.

\section{Acknowledgments}
C.V. and D.R were partially supported by the Belgian Network DYSCO (Dynamical Systems, Control, and Optimization), funded by the Interuniversity Attraction Poles Programme, initiated by the Belgian State, Science Policy Office.
B.E.S. was partially supported by the Department of Energy CSGF (grant number DE-FG02-97ER25308).
I.G.K. was partially supported by the Department of Energy.

\bibliographystyle{model2-names}
\bibliography{coarse_steady_states_and_delayed_initialization}







\end{document}